\documentclass[aps,showpacs,preprint]{revtex4}
\usepackage{latexsym}
\usepackage{epsfig}
\usepackage{amssymb}

\usepackage{graphicx}

\newcommand{\lp}{\left(}
\newcommand{\rp}{\right)}
\newcommand{\lb}{\left[}
\newcommand{\rb}{\right]}

\newcommand{\ba}{\begin{eqnarray}}
\newcommand{\ea}{\end{eqnarray}}
\newcommand{\be}{\begin{equation}}
\newcommand{\ee}{\end{equation}}

\newcommand{\ka}{\kappa}

\newcommand{\R}{\mathcal{R}}

\begin{document}

\title{Hybrid modified gravity unifying local tests, galactic dynamics and late-time cosmic acceleration\footnote{
This essay received an honorable mention in the 2013 Essay Competition of the Gravity Research 
Foundation}}

\author{Salvatore Capozziello$^{1,1a}$}\email{capozzie@na.infn.it}
\author{Tiberiu Harko$^2$}\email{t.harko@ucl.ac.uk}
\author{Francisco S.N.~Lobo$^{3}$}\email{flobo@cii.fc.ul.pt}
\author{Gonzalo J. Olmo$^{4}$}\email{gonzalo.olmo@csic.es}

\affiliation{$^1$Dipartimento di Fisica, Universit\`{a}
di Napoli "Federico II" and  $^{1a}$INFN Sez. di Napoli, Compl.
Univ. di Monte S. Angelo, Edificio G, Via Cinthia, I-80126,
Napoli, Italy
\\
email: capozzie@na.infn.it
}

\affiliation{$^2$Department of Mathematics, University College London, Gower Street, \\
London, WC1E 6BT, United Kingdom \\
email: t.harko@ucl.ac.uk
}

\affiliation{$^3$Centro de Astronomia e Astrof\'{\i}sica da
Universidade de Lisboa, Campo Grande, Ed. C8 1749-016 Lisboa,
Portugal\\
email: flobo@cii.fc.ul.pt
}

\affiliation{$^4$Departamento de F\'{i}sica Te\'{o}rica and IFIC,
Centro Mixto Universidad de Valencia - CSIC. Universidad de
Valencia, Burjassot-46100, Valencia, Spain\\
email: gonzalo.olmo@csic.es
}

\begin{abstract}

The non-equivalence between the metric and Palatini formalisms of $f(R)$ gravity is an intriguing feature of these theories. However, in the recently proposed hybrid metric-Palatini gravity, consisting of the superposition of the metric Einstein-Hilbert Lagrangian with an $f(\cal R)$ term constructed \`{a} la Palatini, the ``true'' gravitational field is described by the interpolation of these two non-equivalent approaches. The theory predicts the existence of a light long-range scalar field, which passes the local constraints and affects the galactic and cosmological dynamics. Thus, the theory opens new possibilities for a unified approach, in the same theoretical framework, to the problems of dark energy and dark matter, without distinguishing a priori matter and geometric sources, but taking their dynamics into account under the same standard.\\

{\bf keywords:} modified gravity, late-time cosmic acceleration, dark matter, Solar System tests

\end{abstract}


\date{\today}

\maketitle


There has been considerable interest in modifications of the
geometric part of  Einstein's field equations, mainly motivated
by the late-time cosmic acceleration and dark matter issues
\cite{expansion1,expansion2, expansion3, expansion4, expansion5, expansion6,expansion7}. In particular, gravitational actions consisting
of more general combinations of curvature invariants than the pure
Einstein-Hilbert term have been investigated extensively
\cite{fRgravity1,fRgravity2, fRgravity3, fRgravity4, fRgravity5, fRgravity6, fRgravity7, fRgravity8, fRgravity9}. Einstein though was more satisfied with the
geometric part of the equations, and has been quoted to say that
while the ``left-hand side is carved of marble, the right-hand
side is made of straw.'' However, in generalized gravity theories,
the problem of coupling matter to gravity is often reduced to the
question of which frame matter resides in with respect to gravity.
The matter Lagrangian and the corresponding stress-energy tensor
are defined in the usual way, but the metric that matter couples
to may be related by a conformal, or more generally a disformal,
transformation to the gravitational metric. Apart from
specific non-conservation terms in the continuity equations, the
structure of the theory is retained.

In this context, a new class of $f(\R)$ gravity theories, denoted
C-theories \cite{Amendola:2010bk}, were considered. In the latter,
the connection was related to the conformally scaled metric $\hat
g_{\mu\nu}=C(\R)g_{\mu\nu}$ with a scaling dependence on the
scalar curvature $\R$. It was shown that the Einstein and Palatini
gravities were obtained as special limits, and in addition to
this, C-theories include completely new physically distinct
gravity theories even when $f(\R)=\R$. With nonlinear $f(\R)$,
C-theories interpolate and extrapolate the Einstein and Palatini
cases and avoid some of their conceptual and observational
problems. In an earlier work
\cite{Flanagan:2003iw}, the known equivalence between higher order
gravity theories and scalar tensor theories was generalized to a
new class of theories. More specifically, in the context of the
Palatini formalism, where the metric and connection are treated as
independent variables (see \cite{Olmo:2011uz} for a recent
review), the Lagrangian density was generalized to a function of
the Ricci scalar computed from the metric, and a second Ricci
scalar computed from the connection. It was shown that these
theories can be written as tensor-multi-scalar theories with two
scalar fields.

More radically, one may modify the response of matter to gravity
by defining an action, which depends nonlinearly upon the matter
Lagrangian \cite{Harko:2010mv,Bertolami:2007gv,Bertolami:2007vu,Bertolami:2008ab,BPHL,HKL}, or its trace
\cite{Harko:2011kv, Haghani:2013oma, Odintsov:2013iba}. Generally, the motion is non-geodesic, and
takes place in the presence of an extra force orthogonal to the
four-velocity \cite{Harko:2012ve}. In fact, in these cases the
motion of matter is typically altered already in flat Minkowski
space, and one may expect instabilities due to new nonlinear
interactions within the matter sector.

A natural way to obtain
solely gravitational modifications of the behaviour of matter
emerges in the Palatini formulation of extended gravity actions.
There the relation between the independent connection and the
metric turns out to depend upon the trace of the matter
stress-energy tensor in such a way that the field equations
effectively feature extra terms given by the matter content.
However, since the extra terms are fourth order in derivatives,
the theory is problematic both at the theoretical and
phenomenological levels \cite{Koivisto:2005yc}.

In the present paper, we propose a simple generalization of the
models resulting from modified gravity actions within the Palatini
formalism, where such problems are absent. An equivalent
formulation is presented as a particular one-parameter class of
scalar-tensor theories. In these theories the scalar field
mediates the new type of coupling to the matter. By a constraint
that is an identity within this class of theories, the scalar
field can be algebraically eliminated in terms of the matter
content and the Ricci curvature in the field equations. Generally
the field equations then feature fourth order derivatives of both the
metric and the matter fields. This provides a consistent and
covariant means to modify the appearance of matter fields in the
field equations, without affecting the conservation laws. This
also opens a new perspective on the $f(R)$ theories, which are the
special limits of the one-parameter class of theories where the
scalar field depends solely on the stress energy trace (Palatini
version) or solely on the Ricci curvature (metric version).

More specifically, consider a one-parameter class of scalar-tensor
theories in which the scalar field is given as an algebraic
function of the trace of the matter fields and the scalar
curvature \cite{Koivisto:2009jn}: \be \label{st_action} S = \int
d^D x \sqrt{-g}\left[\frac{1}{2}\phi R -
\frac{D-1}{2(D-2)\lp\Omega_A-\phi\rp}(\partial\phi)^2 -
V(\phi)\right], \ee
The theories can be parameterized by the
constant $\Omega_A$. The limiting values $\Omega_A=0$ and
$\Omega_A \rightarrow \infty$ correspond to scalar-tensor theories with the Brans-Dicke parameter
 $\omega=-(D-1)/(D-2)$ and $\omega=0$. These limits reduce to
$f(R)$ gravity in the Palatini and the metric formalism,
respectively.

This sheds further light on the nature of $f(R)$ gravity theories.
They can be understood as the -- quite special -- limits of the
action (\ref{st_action}). For any value of $\Omega_A$, the field
can be algebraically eliminated in terms of $R$ and $T$ from the
equations of motion. For any finite value of $\Omega_A$, its value
depends both on the matter and curvature. In particular,
$\Omega_A=1$ corresponds to the case where the deviation from the
Einstein theory is given directly by the deviation from the trace
equation $X=-\kappa^2 T + (D/2-1)R$. In the limit
$\Omega_A \rightarrow \infty$ the propagating mode is given solely
by the curvature, $\phi(R,T) \rightarrow \phi(R)$, and in the
limit $\Omega_A\rightarrow 0$ solely the matter fields $\phi(R,T)
\rightarrow \phi(T)$.  
In the general case the field equations
are fourth order both in the matter and in the metric derivatives.

More specifically, the intermediate theory with $\Omega_A=1$,
corresponds to the hybrid metric-Palatini gravity theory proposed
in \cite{Harko:2011nh}, where the action is given by
\be
\label{action} S= \int d^D x \sqrt{-g} \lb R + f(\R) + 2\ka^2
\mathcal{L}_m \rb\,. \ee
which, in $D=4$, can be recast into a scalar-tensor representation
\cite{Harko:2011nh,Capozziello:2012ny,Capozziello:2012hr} given by
the action
\begin{equation} \label{scalar2}
S= \frac{1}{2\kappa^2}\int d^4 x \sqrt{-g} \left[ (1+\phi)R
+\frac{3}{2\phi}\nabla _\mu \phi \nabla ^\mu \phi -V(\phi)
\right] +S_m \,,
\end{equation}
where $S_m$ is the matter action, $\kappa ^2=8\pi G/c^3$, and
$V(\phi )$ is the scalar field potential. Note that the
gravitational theory given by Eq.~(\ref{scalar2}) is similar to a Brans-Dicke scalar-tensor action with parameter $w=-3/2$, but differs in the coupling to
curvature.

The variation of this action with respect to the metric tensor
provides the field equations
\begin{eqnarray}\label{einstein_phi}
(1+\phi)G_{\mu\nu}=\kappa^2T_{\mu\nu} + \nabla_\mu\nabla_\nu\phi -
\Box\phi\/g_{\mu\nu}
 -\frac{3}{2\phi}\nabla_\mu\phi \nabla_\nu\phi +
\frac{3}{4\phi}\nabla_\lambda\phi\nabla^\lambda\phi g_{\mu \nu}-
\frac{1}{2}Vg_{\mu\nu},
\end{eqnarray}
where $T_{\mu \nu}$ is the matter stress-energy tensor. The scalar
field $\phi$ is governed by the second-order evolution equation
\begin{equation}\label{eq:evol-phi}
-\Box\phi+\frac{1}{2\phi}\nabla _\mu \phi \nabla ^\mu
\phi+\frac{1}{3}\phi\left[2V-(1+\phi)\frac{dV}{d\phi}\right]=
\frac{\phi\kappa^2}{3}T\,,
\end{equation}
which is an effective Klein-Gordon equation
\cite{Harko:2011nh,Capozziello:2012ny}.

In the weak field limit and far from the sources, the scalar field
behaves as $\phi(r) \approx \phi_0 + ( 2G\phi_0 M /3r)
e^{-m_\varphi r}$; the effective mass is defined as $m_\varphi^2
\equiv \left. (2V-V_{\phi}-\phi(1+\phi)V_{\phi\phi})/3\right|
_{\phi=\phi_0}$, where $\phi_0$ is the amplitude of the background
value. The metric perturbations yield
\begin{eqnarray}
h_{00}^{(2)}(r)= \frac{2G_{\rm eff} M}{r} +\frac{V_0}{1+\phi_0}
\frac{r^2}{6}\,, \qquad h_{ij}^{(2)}(r)= \left(\frac{2\gamma
G_{\rm eff} M}{r}-\frac{V_0}{1+\phi_0}\frac{r^2}
{6}\right)\delta_{ij} \label{cor3}\ ,
\end{eqnarray}
where the effective Newton constant $G_{\rm eff}$ and the
post-Newtonian parameter $\gamma$ are defined as
\begin{eqnarray}
G_{\rm eff}\equiv
\frac{G}{1+\phi_0}\left[1-\left(\phi_0/3\right)e^{-m_\varphi
r}\right]\,, \qquad \gamma \equiv
\frac{1+\left(\phi_0/3\right)e^{-m_\varphi
r}}{1-\left(\phi_0/3\right)e^{- m_\varphi r}} \,.
\end{eqnarray}

As is clear from the above expressions, the coupling of the scalar
field to the local system depends on $\phi_0$. If $\phi_0 \ll 1$,
then $G_{\rm eff}\approx G$ and $\gamma\approx 1$ regardless of
the value of $m_\varphi^2$. This contrasts with the result
obtained in the metric version of $f(R)$ theories, and,  as long as $\phi_0$ is sufficiently small, allows to
pass the Solar System tests, even if the scalar field is very light.

In the modified cosmological dynamics, consider the spatially flat
Friedman-Robertson-Walker (FRW) metric $ds^2=-dt^2+a^2(t) d{\bf
x}^2$, where $a(t)$ is the scale factor. Thus, the modified
Friedmann equations take the form
\begin{eqnarray}
3H^2&=& \frac{1}{1+\phi }\left[\kappa^2\rho
+\frac{V}{2}-3\dot{\phi}\left(H+\frac{\dot{\phi}}
{4\phi}\right)\right] \,,\label{field1} \\
2\dot{H}&=&\frac{1}{1+\phi }\left[
-\kappa^2(\rho+P)+H\dot{\phi}+\frac{3}
{2}\frac{\dot{\phi}^2}{\phi}-\ddot{\phi}\right] \,\label{field2}
\end{eqnarray}
respectively.

The scalar field equation (\ref{eq:evol-phi}) becomes
\begin{equation}
\ddot{\phi}+3H\dot{\phi}-\frac{\dot{\phi}^2}{2\phi}+\frac{\phi}{3}
[2V-(1+\phi)V_\phi]=-\frac{\phi\kappa^2}{3}(\rho-3P) .  \label{3}
\end{equation}

As a first approach, consider a model that arises by demanding
that matter and curvature satisfy the same relation as in GR.
Taking
\begin{equation} \label{pot1}
V(\phi)=V_0+V_1\phi^2\,,
\end{equation}
the trace equation automatically implies $R=-\kappa^2T+2V_0$
\cite{Harko:2011nh,Capozziello:2012ny}. As $T\to 0$ with the
cosmic expansion, this model naturally evolves into a de Sitter
phase,  which requires $V_0\sim \Lambda$ for consistency with
observations. If $V_1$ is positive, the de Sitter regime
represents the minimum of the potential.  The effective mass for
local experiments, $m_\varphi^2=2(V_0-2 V_1 \phi)/3$, is then
positive and small as long as $\phi<V_0/V_1$. For sufficiently
large $V_1$ one can make the field amplitude small enough to be in
agreement with Solar System tests. It is interesting that the
exact de Sitter solution is compatible with dynamics of the scalar
field in this model.

Relative to the galactic dynamics, a generalized virial theorem,
in the hybrid metric-Palatini gravity, was extensively analyzed
\cite{Capozziello:2012qt}. More specifically, taking into account
the relativistic collisionless Boltzmann equation, it was shown
that the supplementary geometric terms in the gravitational field
equations provide an effective contribution to the gravitational
potential energy. The total virial mass is
proportional to the effective mass associated with the new terms
generated by the effective scalar field, and the baryonic mass.
This shows that the geometric origin in the generalized virial
theorem may account for the well-known virial  mass
discrepancy in clusters of galaxies. In addition to this,
astrophysical applications of the model where explored, and it was
shown that the model predicts that the effective  mass associated to the
scalar field, and its effects, extend beyond the virial radius of
the clusters of galaxies. In the context of the galaxy cluster
velocity dispersion profiles predicted by the hybrid
metric-Palatini model, the generalized virial theorem can be an
efficient tool in observationally testing the viability of this
class of generalized gravity models. Thus, hybrid metric-Palatini
gravity provides an effective alternative to the dark matter
paradigm of present day cosmology and astrophysics.

In a monistic view of Physics, one would expect Nature to make somehow a choice between the two distinct possibilities offered by metric and Palatini formalisms. We have shown, however, that a theory consistent with observations and combining elements of these two standards is possible. Hence gravity admits a diffuse formulation where mixed features of   both formalisms allow to successfully address large classes of phenomena.

\section*{Acknowledgments}

We thank Tomi Koivisto for enlightening discussions, a careful reading of the manuscript and
for very helpful comments. FSNL acknowledges financial support of
the Funda\c{c}\~{a}o para a Ci\^{e}ncia e Tecnologia through the
grants CERN/FP/123615/2011 and CERN/FP/123618/2011. G.J.O. has been supported by the Spanish grant FIS2011-29813-C02-02,  and the JAE-doc program of the Spanish Research Council (CSIC).



\end{document}